\documentclass[twocolumn,10pt]{article}
\usepackage{amsmath,amssymb,epsfig,graphicx}
\newcommand{\be}{\begin{equation}}
\newcommand{\ee}{\end{equation}}

\title{Load distribution in small world networks}

\author{M. di Bernardo, F. Garofalo, S. Manfredi, F. Sorrentino
\thanks{Corresponding author. Email: {\small fsorrent@unina.it}}
\thanks{The authors are listed in alphabetical order}
\\
Faculty of Engineering, University Federico II, Naples, Italy}

\date{}

\begin{document}
\maketitle
%

%
\begin{abstract}
In this paper we introduce a new model of data packet transport,
based on a stochastic approach 
with the aim of characterizing the load distribution on complex
networks. Moreover we analyze the load standard deviation as an
index of uniformity of the distribution of packets within the
network, to characterize the effects of the network topology. We
measure such index on the model proposed by Watts and Strogatz as
the redirection probability is increased. We find that the
uniformity of the load spread is maximized in the intermediate
region, at which the small world effect is observed and both
global and local efficiency are high. Moreover we analyze the
relationship between load centrality and degree centrality as an
approximate measure of the load at the edges. Analogous results
are obtained for the load variance computed at the edges as well
as at the vertices.


\end{abstract}
\section{Introduction}

Congestion in real networks is a complex phenomenon that depends
on a large number of variables. Here we are interested in
understanding how the underlying structure of the network can
itself have an influence on congestion. The network is supposed to
be crossed by information units or packets. We will consider a
packet transport model on a given graph and will study the effects
of changes in the graph topology on the packets distribution. With
the aim of understanding the effects of the network topology on
congestion, we will keep constant other variables as the rate at
which packets are generated, the distribution of sources and
destinations within the network, and the routing algorithm to go
from the former to the latter. We will show that each vertex can
draw toward itself an higher or lower flow of packets, according
to its own \emph{position} within the graph.

\section{The network as a Markov Chain}

In the recent literature on networks several dynamic packet
transport models were analyzed with the aim of describing
realistic communication over the network  \cite{ARROW04}
\cite{So:Va} \cite{Oh:Sa}\cite{Manfredi2} \cite{Manfredi1}. On the
other hand in \cite{korea} \cite{korea3} some static parameters
were introduced in order to characterise the load distribution
over the network without considering explicitly its dynamics.

We will suggest here an alternative and intermediate packet
transport model based on a stochastic approach.

We consider a network with $N$ vertices. We make the hypothesis
that the origin and the destination of each packet are chosen with
uniform probability within all the vertices in the network. We
suppose that at each time step, each vertex generates $N-1$
packets, each one addressed to every other node in the network so
that the total number of packets generated at each time step is
equal to $N(N-1)$.

Then packets are supposed to be routed from origin to destination
along the geodesic, i.e. according to a minimum distance
criterion. We suppose that packets are delivered according to a
dynamic process that takes into account the actual time needed for
packets to travel across the network. We consider the case of a
parallel process, in which at each time unit all packets move
simultaneously, every one crossing one edge. We assume that the
time in which a packet goes through exactly one edge is equal to
one time unit. 
Specifically we assume that at each time step:
\begin{enumerate}
    \item Packets are generated at every vertex.
    \item Packets in a given node
    but not delivered yet to their final destinations, are routed to
    the neighbors of that node that are nearest to the destination.
    \item Packets delivered to their final destination are removed from the network.
\end{enumerate}

In \cite{korea}, the load at a given vertex $v$, $\ell(v)$, was
defined as the number of shortest paths between pairs of nodes
crossing it. Analogously we will define here the load at the edges
$\ell(e)$, as the number of shortest paths crossing a given edge.

Let $J(i)$ be the subgraph of the vertices $j$ adjacent to $i$. As
a packet is crossing $i$, it can take one of the directed edges
$e_{ij}$, $j\in J(i)$. We assume that each edge $e_{ij}$ in the
graph is crossed with frequency proportional to its load $l(e)$.
Consistently with that assumption, we will estimate the
conditional probability of a packet starting from $i$, going
through one of the edges $e_{ij}$, as:
\begin{equation}
\label{eq:pij}
p(e_{ij}/i)=\frac{\ell(e_{ij})}{\sum_{j}\ell(e_{ij})}.
\end{equation}

Using (\ref{eq:pij}), we can now associate to the network of
interest a Markov Chain; each node being represented as a state of
the chain with the probability of going from a state $i$ to a
state $j$ being given by $p(e_{ij}/i)$ for $j\in J(i)$.

The Markov Chain represents a dynamic model describing the
evolution of the packets distribution in time. 
We can obtain numerically the limit distribution $\Delta$, i.e.
the probability of being asymptotically in a certain state/node of
the Markov Chain. We wish to emphasize that $\Delta$ can be used
as an alternative measure of the load at vertices. In particular,
we found $\Delta$ to differ from the previous estimation of the
load at vertices \cite{korea}.

We can give a physical explanation of that, according to the
parallel nature of the packets exchange over the network.

\begin{figure}[tbhp]
\begin{center}
\epsfig{width=0.5\textwidth, file=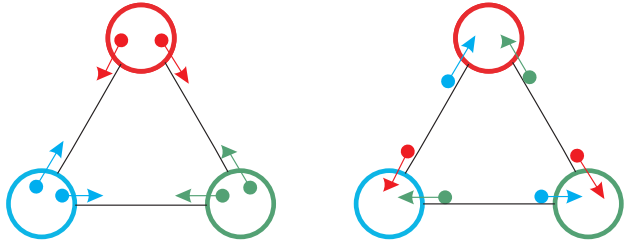} \caption{\label{fig:espl}
\small A parallel model of packets exchange in a sample network of
three nodes. We represent with the same color packets
\emph{generated} at a given vertex. In the left panel we represent
the distribution of packets after their generation. 
It is worth noticing that the distribution does not change if we
look at it after packets have been delivered. 
As we can see, packets just having been generated and the ones
delivered to destination are never present simultaneously in the
same node at any given time. }
\end{center}
\end{figure}

Fig.1 shows the load distribution of packets in a representative
network on which packets are exchanged in a parallel way. We
observe that for each packet travelling from a vertex $i$ to a
vertex $j$, there is another one travelling from $j$ to $i$. This
implies that packets just having been generated and the ones
delivered to destination are never present simultaneously in the
same node at any given time.

Let's define $\Lambda_v$ the total load at the vertices in the
network, computed as the sum of the loads over all the vertices
within the network, i.e. $\Lambda_v=\sum_{v}\ell(v)$. We have
that:

\begin{equation}
\label{eq:totload} \Lambda_v=d\cdot N(N-1),
\end{equation}
where $d$ is the characteristic path length associated to the
graph. From (\ref{eq:totload}), we have that the average load at
vertices is given by $\overline{l}=d\cdot(N-1)$. We observe that
the average load is proportional to the characteristic path length
of the graph: shortening mean distances between vertices implies
that packets travel from origin to destination in fewer steps.
Hence, a load reduction is observed on the whole network.

Moreover, it is worth mentioning here that expression
(\ref{eq:totload}) is identical to Little's Law \cite{LittleBOOK}
in queueing theory, i.e.
\begin{equation}
\mathcal{V}=\tau\cdot N(N-1),
\end{equation}
where the total number of packets in the network $\mathcal{V}$ is
replaced with $\Lambda_v$ and  $\tau$, the mean time spent by a
packet going from origin to destination in Little's law, is
replaced with $d$, under the hypothesis that each edge is crossed
within the same amount of time.

Note that, according to the newly introduced load definition, we
find that, contrary to what presented in \cite{korea}, the total
load at the vertices is equal to the total load at the edges
$\Lambda_e=\sum_{e}l(e)$, i.e. $\Lambda_e=\Lambda_v$. This is
expected as in our model, the total amount of packets within the
network at any time has to be the same.


\section{Uniformity of load distribution in small world networks}

We will try now to assess whether the network topology can itself
cause a more or less uniform
distribution of packets in the network. 
In \cite{Ar:dB:So} the effects of variations of load distribution
in random and scale free networks were analyzed. The former are
characterized by a Poisson degree distribution while the second
have a degree distribution of the form $P(k) \sim k^{-\gamma}$.
From a communication point of view, it would be desirable to have
a uniform load distribution in order to exploit evenly the network
resources (nodes and edges).
In particular we will try to understand how uniform over a given
network the load distribution is, because of its topology.


Specifically, an high variance of that distribution should
indicate an unfair use of the network, and could therefore point
out a possible cause of congestion.

In \cite{Gu:Gu03}, the maximum load $\ell_{max}$ was proposed as a
main index for characterizing the network structure and an high
value of this parameter was claimed responsible for congestion.
Nevertheless, this parameter does not describe the whole structure
of the graph, giving information only on the load level at the
hub. For that reason here we will evaluate the load standard
deviation as defined in \cite{Ar:dB:So}, that refers to the
network as a whole. In order to make it insensitive to the average
values of $\ell$, we will evaluate $\tilde{\sigma}(\ell)$, the
standard deviation of the load
appropriately normalized with respect to its mean. 

The standard deviation of the load distribution $\sigma(\ell)$ was
shown to have an effect on network performances in terms of
throughput and delivery time. Namely an high value of the load
variance, that is typical of scale-free networks was seen to be
able to worsen strongly the network performances.

We analyze here the variation of load distribution both at the
edges and at the vertices due to changes in the underlying
topology.

A common feature of most real networks is that they are small
world, i.e. the average distance between randomly chosen nodes is
generally low and increases only logarithmically with the network
size. Moreover real networks show high clustering or transitivity,
that is, high probability that two randomly selected neighbors of
a node are also neighbors of each other. Watts and Strogatz (WS)
\cite{Wa:St98} showed that, starting from a regular nearest
neighbor circle, a few random redirection were sufficient to get
the small world effect, without compromising the clustering.

We will repeat the experiment in \cite{Wa:St98}, measuring also
the normalized standard deviation of the vertices load,
$\tilde{\sigma}(\ell(v))$ and the edges load
$\tilde{\sigma}(\ell(e))$.

\begin{figure}[tbp]
\begin{center}
\epsfig{width=0.5\textwidth, file=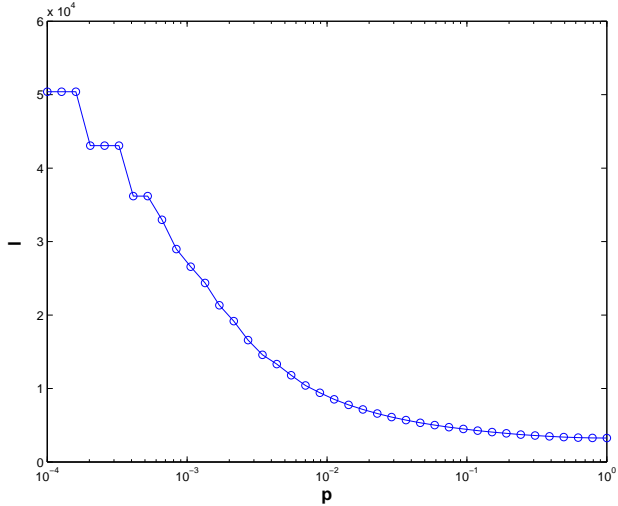}
\caption{\label{fig:art2mille} \small The WS model. A regular
lattice with 1000 vertices and 10 links per vertex is rewired with
probability $p$. Average vertices load
$\overline{\ell}=d\cdot(N-1)$ is reported as a function of $p$.
The results are obtained over 100 diverse experiments.}
\end{center}
\end{figure}

\begin{figure}[bhp]
\begin{center}
\epsfig{width=0.5\textwidth, file=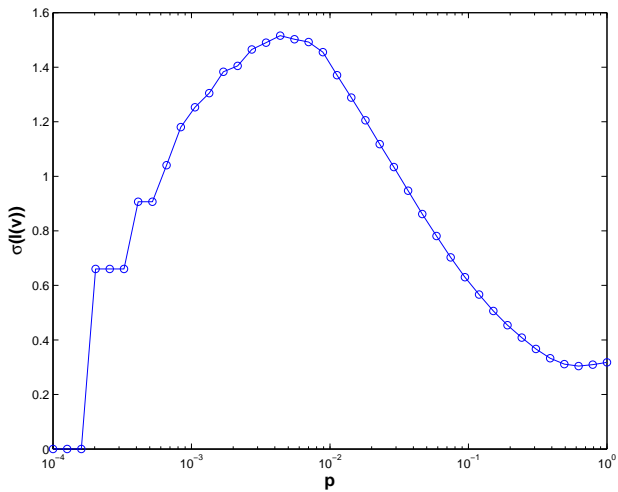}
\caption{\label{fig:art1} \small Normalized standard deviation of
vertices load $\tilde{\sigma}(\ell(v))$ is reported as varying the
redirection probability, under the same hypothesis of figure 2}
\end{center}
\end{figure}

In Fig. 2 the average load in the graph is shown as the
redirection probability $p$ is varied. It can be observed that the
average load behaves as the characteristic path length of the
graph, decreasing as $p$ increases.

The normalized standard deviation of the vertices load is shown in
Fig. 3. Here, we see that such quantity increases at first,
then over a certain threshold value of the redirection probability, 
it is progressively reduced. In fact, as the redirected edges are
few, the global reduction of the characteristic path length is
achieved through the exploitation of the same shortcuts and the
vertices situated at their endpoints.

A further increase of $p$ (i.e. of the number of redirected edges)
causes an increase of the number of different paths linking pairs
of nodes in fewer steps. Thus, for higher $p$, an higher load
uniformity (lower $\tilde{\sigma}(\ell(v)$) is observed. In other
words we observe that, starting from a regular configuration, and
adding progressively some disorder to it, the parameter increases
at first and then decreases after a critical point. So we have
that at intermediate $p$, at which local and global efficiency are
both maximized \cite{La:Ma02ec}, we get the least uniform load
distribution between vertices. In Fig. 4 we report the median (in
blue) of the load distribution with the 25th and 75th percentile
(in red), showing that the interquartile range of the sample is
larger for intermediate values of $p$.

We can argue thus that the small-world effect, i.e. the reduction
of the average distance between vertices caused by the addition of
a relatively small number of shortcuts, can be associated with an 
unfair spread of the load distribution across the network. This
would cause an unfair exploitation of some of the network
resources, making some areas of the network more likely to be
interested by higher traffic and therefore by congestion
phenomena.

As shown in Fig. 5, similar results are obtained when the load
standard deviation at the edges (rather than at the vertices) is
computed.

\begin{figure}[thp]
\begin{center}
\epsfig{width=0.5\textwidth, file=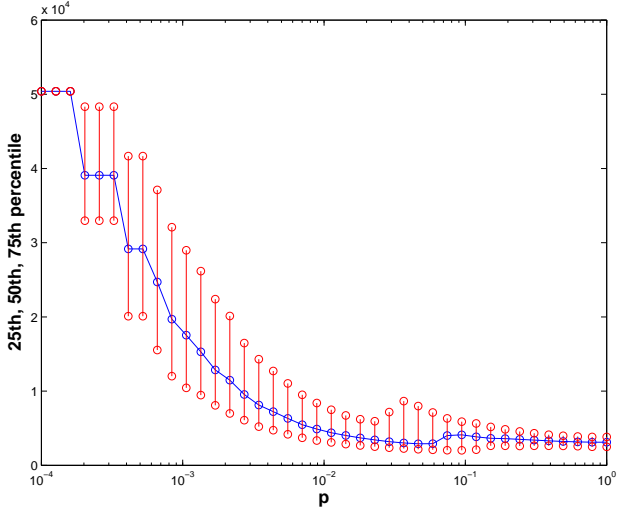}
\caption{\label{fig:art4} \small The median of load distribution
is reported as varying the redirection probability (blue line),
under the same hypothesis of figure 2. Upper and lower red circles
represent the 25th and the 75th percentile, with the length of the
red segments indicating the interquartile range.}
\end{center}
\end{figure}

\begin{figure}[thp]
\begin{center}
\epsfig{width=0.5\textwidth, file=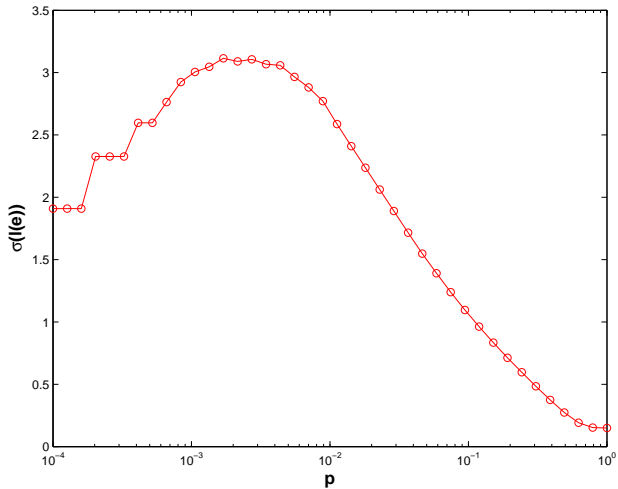}
\caption{\label{fig:art3mille} \small Normalized STD of edges load
$\tilde{\sigma}(\ell(e))$ is reported as varying the redirection
probability, under the same hypothesis of figure 2}
\end{center}
\end{figure}

\section{Load Distribution And Degree}

As discussed above, by adding shortcuts at a given vertex, a load
increase is observed at the node as well as at the edges leaving
from it. So we can notice two different effects of rewiring links
on the load at the edges: (i) a bigger amount of packets is routed
by the starting node through the outgoing edges; (ii) the load is
shared between an higher number of edges. Therefore, the load
distribution at the edges is influenced by both the load and the
degree at the vertices which they leave from.

From this point of view, it is worth understanding the
relationship between the two main measures of point centrality
\cite{SNAbook}, the one based on the degree and the
other on the \emph{betweenness}. 

A first contribution in that sense was given in \cite{Be03}, where
it was shown that, in the case of a power law degree distribution,
the load scaling were characterized by a function of the degree
$k$:
\begin{equation}
\label{eq:BE} P(\ell)\propto k^{\eta},
\end{equation}
This means that nodes with more incident edges are the ones that
draw on themselves more load, according to an assigned power of
the degree.


Here, to define a more general link between load and degree, we
introduce a new index, the load-degree ratio $r(i)$ at each vertex
i, defined as
\begin{equation}
r(i)= \frac{\ell(i)}{k(i)} = \frac{\sum_{j \in J(i)}
\ell(e_{ij})}{k(i)}.
\end{equation}
This parameter measures the number of packets at each vertex, with
respect to the number of incident links and moreover it is an
average value of the load at the edges incoming or outgoing from
$i$.


We observe that, when the redirection procedure is applied, the
degree of some vertices increases (as new edges are connected to
them). Correspondingly a significant increase of the load at
vertices is also detected. Therefore, the load-degree ratio of
those vertices increases as well. On average this means that the
load at the edges leaving from overloaded vertices becomes larger.
This explains the behavior of $\tilde{\sigma}(\ell(e))$ in Fig. 5
below the threshold value of $p$.

On the contrary, for further variations of the redirection
probability, as the network approaches a random configuration,
$\tilde{\sigma}(\ell(e))$ decreases similarly to what observed for
$\tilde{\sigma}(\ell(v))$.

\section{CONCLUSION}

In this work we have introduced a new load definition more
suitable in the case of a dynamical process in which packets
travel across the network in parallel. This new formulation
differs from previous ones as it takes into account in the
computation alternatively the contribution of the packets yet
generated or of the ones delivered to destination; in fact in a
parallel process packets outgoing from a given vertex move at the
same time as the incoming ones arrive.


Moreover we have introduced two new indices with the aim of
evaluating the role of the network topology in influencing the
packets distribution within the network, the standard deviation of
the load at the vertices $\tilde{\sigma}(\ell(v))$, and at the
edges $\tilde{\sigma}(\ell(e))$. We have reworked through the WS
experiment, computing the new quantities. We have noticed that as
the small-world effect can be obtained by the addition of a
relatively low number of shortcuts, this is made at the expenses
of the uniformity of the load distribution within the network.


\end{document}